\documentclass[twocolumn,astrosymb, times]{aastex701}
\usepackage{rotating}
\usepackage{booktabs}
\usepackage{textcomp, gensymb}
\usepackage{float}
\usepackage{tabularx}
\usepackage{amsmath}
\usepackage{textgreek}

\shortauthors{P\'erez Paolino et al.}

\begin{document}

\title{Rotational Modulation and Long-Term Variability of Magnetic Fields in T-Tauri Stars with IGRINS }
\correspondingauthor{Facundo P\'erez Paolino}
\author[0000-0002-4128-7867, gname=Facundo, sname=P\'erez Paolino]{Facundo P\'erez Paolino}
\affiliation{Department of Astronomy, California Institute of Technology, 1216 East California Blvd, Pasadena, CA 91125, USA}
\email[show]{fperezpa@caltech.edu}
\author{Lynne A. Hillenbrand}
\affiliation{Department of Astronomy, California Institute of Technology, 1216 East California Blvd, Pasadena, CA 91125, USA}
\email{lah@astro.caltech.edu}

\author[0000-0001-8642-5867]{Jeffrey S. Bary}
\email{jbary@colgate.edu}
\affiliation{Colgate University, 13 Oak Drive, Hamilton, NY 13346, USA}

\begin{abstract}
Magnetic fields play a central role in the evolution of pre-main-sequence (PMS) stars, yet direct observational constraints on their variability over rotational and multi-year timescales remain scarce. We investigate the temporal behavior of surface magnetic fields in a sample of nine PMS stars observed with the Immersion GRating INfrared Spectrometer (IGRINS), using 489 high-resolution near-infrared spectra drawn from the Raw and Reduced IGRINS Spectral Archive. We fit each epoch with magnetic synthetic spectra to derive the mean surface magnetic field strength $\langle Bf \rangle$ and detect correlated magnetic and thermal variability in six of the nine stars while being able to recover the known stellar rotation period in at least one observing season for all six. We find that not only the mean magnetic field strength and effective temperature evolve on year-long baselines, but so does the amplitude of the rotational modulation signal (which in some cases weakens or disappears entirely). This behavior indicates that magnetic variability is driven by both changes in the total magnetic flux and the spatial distribution and contrast of surface magnetic inhomogeneities. For two stars in the sample with starspot measurements, we find that the magnetic filling factors are systematically larger than those inferred from temperature, implying that magnetic regions extend beyond the coolest spotted areas and occupy a broader fraction of the stellar surface (i.e., plages). These results provide direct evidence that PMS magnetic variability is structured, rotationally modulated, and evolves on year timescales.

\end{abstract}


\section{Introduction}
Magnetic fields influence nearly every aspect of a star’s life, yet our understanding of their origin, structure, and variability remains incomplete \citep{Basri2021}. Observations of the solar dynamo reveal a highly variable magnetic topology, with variability on short and long timescales \citep{Charbonneau2010, Brun2017}. Over the last three decades, growing bodies of evidence have shown that other cool stars experience similar magnetic activity cycles with multi-year baselines, both in chromospheric diagnostics and in the large-scale magnetic field topology \citep[e.g.,][]{Baliunas1995, Donati2009,See2016, Williamo2022}. Spectropolarimetric techniques such as Zeeman–Doppler Imaging (ZDI) have been used to map large-scale magnetic fields over multiple epochs, revealing topology changes and polarity reversals in stars like 61 Cyg A and τ Boo, analogous to the solar magnetic cycle \citep{See2016, Williamo2022}.

A growing body of evidence points to magnetic variability being a ubiquitous phenomenon in sun-like stars. \citet{Cristofari2025} presented a detailed analysis of the rotational modulation and long-term evolution of the total unsigned field in a sample of M dwarfs observed with SPIRou. Using high-resolution near-infrared spectropolarimetry, they tracked variations in Zeeman broadening and polarization signatures over rotational and multi-year timescales, finding rotationally-modulated magnetic variability in four stars. Their results are consistent with the findings of \citet{PerezPaolino2025separating} that magnetic surface fields in fully and partially convective young pre-main-sequence stars are concentrated in starspots, and should undergo rotational variability similar to that observed in molecular absorption bands \citep{myself}.

Pre-Main-Sequence (PMS) stars are known to host strong surface magnetic fields, with mean strengths of order kilogauss \citep[e.g.,][]{johnskrull_magnetic_2007, Donati2019, Flores2022}, and theoretical models predict substantial magnetic variability as both rotation and internal structure evolve rapidly during this phase \citep{gregory_can_2012}. Yet despite these expectations, direct observational constraints on the temporal evolution and variability of magnetic field strength and topology in PMS stars remain remarkably limited.


Given the importance of magnetic fields in controlling exoplanet evolution and habitability \citep{Airapetian2017}, understanding not only the mean strength but also the temporal variability of stellar magnetic fields is essential. {Stellar magnetism regulates high-energy X-ray and EUV emission (via coronal heating) and stellar winds that govern atmospheric erosion, ion escape, and surface irradiation histories.} Yet most exoplanet studies implicitly assume time-averaged \citep[e.g.,][]{Owen2012, Johnstone2021} or solar calibrated \citep[e.g.,][]{Ribas2005, Fetherolf2025, Tu2015} magnetic prescriptions. This is despite growing evidence that young and magnetically active stars exhibit orders of magnitude variability on rotational and secular timescales.


The paper is structured as follows. In Section~\ref{sec:obs}, we describe the selected sample of archival IGRINS data. In Section~\ref{sec:methods}, we outline our method for measuring magnetic fields from the spectra. In Section~\ref{sec:mag_var}, we present our results on magnetic variability in young stars, and in Section~\ref{sec:corr}, we examine correlations between magnetic and temperature variability. This analysis is extended in Section~\ref{sec:surf}, where we infer the surface coverage of magnetic regions. Finally, in Section~\ref{sec:discussion}, we place our results in context and summarize our conclusions in Section~\ref{sec:conclusion}.

\section{Sample Selection}\label{sec:obs}
\begin{deluxetable*}{lccccccccc}
\tablecaption{
{Median Parameters and Standard Deviations Across All Epochs.}
\label{tab:stars_summary}
}\tablehead{\colhead{Star} & \colhead{$N_{\rm obs}$} & \colhead{$\widetilde{T_{eff}}$} & \colhead{$\widetilde{\log g}$} & \colhead{$\widetilde{v\sin i}$} & \colhead{$\widetilde{\langle Bf \rangle}$} & \colhead{$P_{\rm rot}$} & \colhead{$P_{\rm rot}$ Source}\\ \colhead{} & \colhead{} & \colhead{(K)} & \colhead{(cm\,s$^{-2}$)} & \colhead{(km\,s$^{-1}$)} & \colhead{(kG)} & \colhead{(days)}}
\startdata
AA Tau & 31 & $3825\pm{28}$ & $4.11\pm{0.14}$ & $12.79\pm{0.60}$ & $2.25\pm{0.14}$ & 8.20 & 1\\
CI Tau & 72 & $4051\pm{46}$ & $3.97\pm{0.09}$ & $12.27\pm{1.08}$ & $1.93\pm{0.18}$ & 9.04 & 2\\
GI Tau & 21 & $3778\pm{44}$ & $4.16\pm{0.08}$ & $11.65\pm{1.25}$ & $2.15\pm{0.15}$ & 7.13 & 2\\
IQ Tau & 23 & $3778\pm{29}$ & $4.19\pm{0.15}$ & $13.40\pm{0.47}$ & $1.81\pm{0.11}$ & 6.67 & 2\\
L1551-51 & 88 & $4204\pm{36}$ & $4.55\pm{0.04}$ & $30.33\pm{1.16}$ & $2.60\pm{0.19}$ & 2.43 & 2\\
LkCa 15 & 29 & $4256\pm{33}$ & $4.58\pm{0.07}$ & $13.72\pm{0.01}$ & $1.76\pm{0.13}$ & 5.77 & 1\\
V827 Tau & 73 & $3776\pm{30}$ & $4.22\pm{0.05}$ & $19.68\pm{0.63}$ & $2.61\pm{0.16}$ & 3.76 & 2\\
V830 Tau & 85 & $3889\pm{27}$ & $4.39\pm{0.06}$ & $29.87\pm{0.55}$ & $3.00\pm{0.14}$ & 2.74 & 2\\
V1023 Tau (Hubble~4) & 67 & $3878\pm{14}$ & $3.93\pm{0.06}$ & $16.52\pm{0.51}$ & $2.73\pm{0.05}$ & 1.55 & 2\\
\enddata
\vspace{0.1cm}
(1) \citet{Bouvier07_magnetospheric_accretion-ejection}, (2) \citet{Rebull2020}.

\end{deluxetable*}

The {publicly available} Raw and Reduced IGRINS Spectral Archive (RRISA; \citealt{RRISA}) contains nearly 18,000 telluric corrected, high-resolution near-infrared spectra of stars obtained with the Immersion GRating INfrared Spectrometer\footnote{\url{https://github.com/igrins/plp}} (IGRINS; \citealt{Park2014}). These observations span approximately 2,700 unique sources and were collected at the 2.7-m Harlan J. Smith Telescope at McDonald Observatory, the 4.3-m Lowell Discovery Telescope, and the 8.1-m Gemini South telescope. Operating at a resolving power of $R \approx 45{,}000$ and employing a fixed optical design, IGRINS delivers spectra that are highly uniform in both resolution and wavelength calibration across all observatories, with negligible differences in radial velocity zero points or line-spread functions. Its simultaneous coverage of the H and K bands makes IGRINS particularly well suited for studies of stellar magnetic fields, a capability we exploit extensively in this work.

From RRISA, we selected stars with at least five observations at signal-to-noise ratios $\mathrm{SNR} \geq 75$, yielding a final sample of $9$ sources spanning 489 individual observations. These targets are listed in Table~\ref{tab:stars_summary}, along with stellar rotation periods where available. The observations span more than a decade, from 2014 to 2024, and include several stars with over 50 individual visits distributed across multiple observing campaigns. For these high-coverage targets, the data enable measurements of magnetic variability on both rotational timescales within individual observing runs and multi-year timescales across separate campaigns. 

All of the stars in our sample are well-studied young sun-like stars in the Taurus-Auriga star forming region \citep[e.g.,][]{herczeg2014}. Four of them are actively accreting Classical T Tauri stars, while the remaining 5 stars are non-accreting weak-lined T Tauri Stars. Additionally, three of them are exoplanet candidates (CI Tau; \citet{JohnsKrull16_candidate_young}, V830 Tau; \citet{Damasso2020}, LkCa 15 \citet{Currie19_no_clear}). Part of the 2016-2017 data for CI~Tau was already presented in \cite{sokal2020}, while a majority of the remaining CI~Tau and V830~Tau data have been presented in \cite{Johns-Krull2022AAS}, who used magnetically sensitive Ti~I lines in the K band to determine magnetic field strengths. We re-analyze both of these datasets here.

\section{Methodology}\label{sec:methods}
We follow the approach of \citet{lopez-valdivia_igrins_2023} to extract stellar parameters and surface magnetic field strengths from IGRINS spectra, with a few modifications. Our analysis employs a grid of magnetic stellar atmosphere models from \citet{LopezValdiviaCode}, computed using MOOGSTOKES, which builds on the 1D~LTE radiative transfer code MOOG by incorporating the effects of Zeeman broadening due to a purely radial magnetic field geometry. {For the calculation, underlying synthetic atmospheres are taken from MARCS models \citep{Gustafsson2008} assuming solar metallicity \citep{Grevesse07_solar_chemical}, while the atomic transitions are based on the Vienna Atomic Line Database (VALD; \citealt{Ryabchikova2015}) with improved van der Waals coefficients and oscillator strengths taken from \citet{Flores2019}.} The model grid spans effective temperatures from 3000 to 5000~K, with steps of 100~K over the first 1000~K and 250~K thereafter. Surface gravities cover $3.0 \leq \log g \leq 5.0$ in increments of 0.5 dex, projected rotational velocities range from 2 to 50~km s$^{-1}$ in steps of 2~km s$^{-1}$, and magnetic field strengths span 0 to 4~kG in steps of 0.5~kG. The grid was then linearly interpolated to provide continuous parameter coverage. {We used a microturbulence of 1~km s$^{-1}$, typical of young stars, following \citep{lopez2021}.}

Young PMS stars need additional consideration when compared to main-sequence stars. In the near-infrared, continuum emission from warm dust in the inner disk veils, or reduces, the depth of absorption lines, making them appear weaker. We therefore continuum-normalized our spectra using an asymmetric least-squares algorithm \citep[e.g.,][]{Eilers} before veiling our models following \citet{Hartigan1991}:
\begin{equation}
F_{\lambda,\mathrm{veil}} = \frac{F_{\lambda,\mathrm{phot}} + r_{H/K}}{1 + r_{H/K}},
\end{equation}
\noindent
where $r_{H/K}$ represents the continuum veiling at $H$ or $K$ band.
Fitting was performed at the native sampling resolution to preserve line shape and strength using a Markov Chain Monte Carlo (MCMC) sampler \citep[\texttt{emcee;}][]{Foreman-Mackey2013}, with flat priors and a likelihood function:
\begin{equation}
\ln p = -\frac{1}{2} \sum_n
\left[ \frac{(y_{\mathrm{data,\ i}} - y_{\mathrm{model,\ i}})^2}{\sigma^2_i}
+ \ln(2\pi\sigma_i^2) \right].
\end{equation}

We increase the flux errors reported by the reduction pipeline using a fractional uncertainty $k_{\mathrm{unc}}$ that accounts for small imperfections in the synthetic spectra \citep{PerezPaolino2025separating}. Effective uncertainties are therefore 

\begin{equation}
\sigma_{\mathrm{eff}} = \sqrt{\sigma_{\mathrm{data}}^2 + (y_{\mathrm{model}}k_{\mathrm{unc}})^2},
\end{equation}

\noindent
where $k_{\mathrm{unc}} = 0.01$.

Variations in effective temperature with rotational phase are expected in heavily spotted PMS stars \citep[e.g.,][]{myself,Tang2024}, as are changes in the measured surface magnetic field strength. In contrast, neither surface gravity nor projected rotational velocity is expected to vary on rotational or multi-year timescales. We therefore interpret the scatter in $\log g$ and $v\sin i$ obtained from the initial, fully free fits as arising from statistical uncertainties, rather than physical variability. This is similar to the approach of \citet[e.g.,][]{Cristofari2025} using a template average over all epochs. By fixing these quantities, we are able to isolate smaller, physically meaningful variations in $T_{\mathrm{eff}}$ and $\langle Bf\rangle$ that might otherwise be obscured by stochastic fluctuations in surface gravity or rotational broadening. 

During fitting, we allow the veiling ($r_{H}$, $r_{K}$ to vary, as variable accretion rates can lead to variable veilings that alter line shapes. In practice, all of our observations returned veiling values under 2 ($r_{H}$ and $r_{K}$ $<2$), and veiling is not expected to affect uncertainties in magnetic field strengths significantly \citep{drouglazet_magnetic_2026}.

{Fitting was performed over the spectral regions shown in Figure~\ref{fig:min_vs_max_B}. These include magnetically sensitive lines, such as the Na~\textsc{I} doublet at 2.2062 and 2.2090~$\mu$m, Sc~\textsc{I} lines at 2.2058 and 2.2071~$\mu$m, and the temperature-sensitive Si~\textsc{I} line at 2.2068~$\mu$m. Additional diagnostics include magnetically sensitive Ti~\textsc{I} lines at 2.2217, 2.2239, and 2.2315~$\mu$m, Fe~\textsc{I} lines sensitive to $T_{\mathrm{eff}}$, and the weakly magnetic Ca~\textsc{I} lines at 2.2614, 2.2631, and 2.2657~$\mu$m (along with Fe~\textsc{I} at 2.2626~$\mu$m), which are primarily sensitive to $\log g$. Finally, the 2.2986–2.3150~$\mu$m region includes several CO band heads that strengthen toward lower surface gravity, providing an additional constraint on $\log g$. }CO is also an excellent determinant of vsini, due to its magnetic insensitivity.{ While the K-band is well suited for determining magnetic field strengths, it offers limited constraints on stellar effective temperatures. In \citet{PerezPaolino2025b} we found that combining the $K$-band with the $H$-band during fitting provided effective temperatures that showed small ($\leq$~100~K) differences between single temperature fits and full spotted models that separately measure the spot temperature and coverage. Given the spotted nature of young stars \citep{Cao2022}, we decided to combine the magnetically sensitive 2.200–2.315~$\mu$m window in the $K$ band with a 1.558–1.568~$\mu$m segment in the $H$ band. This $H$-band interval contains several temperature-sensitive lines, notably Fe~\textsc{i} at 1.56259 and 1.56362~$\mu$m and OH at 1.56310 and 1.56317~$\mu$m, which vary inversely with temperature over the 3000–5000~K range \citep{Tang2024}.}

Each spectral model is thus characterized by seven free parameters that are fit simultaneously: $T_{\mathrm eff}$, $\log g$, $v\sin i$, $v_{\mathrm{rad}}$, $\langle Bf\rangle$, $r_H$, and $r_K$. We used this routine on every epoch of selected IGRINS data with 50 walkers started randomly across parameter space ran for 5000 steps each. {In practice, convergence was obtained within ~200 steps, and we discarded the first 2500 steps as burn-in. This is despite the fact that in each run, the limits for the fitting were kept at the full width of the parameter space (e.g., $\log g = 3$–5 dex, $T_{\mathrm{eff}} = 3000$–5000 K).} An example of a corner plot is shown in Figure~\ref{fig:corner}, showing no evidence for any correlations between parameters or e.g., bimodal solutions.

In Table~\ref{tab:stars_summary} we report the median values of $T_{\mathrm{eff}}$, $\log g$, and $v\sin i$ for each star, computed across all available epochs and report these values. To investigate epoch-to-epoch magnetic variability, we then refit each individual spectrum while fixing $\log g$ and $v\sin i$ to their median values, but allowing $T_{\mathrm{eff}}$ and $\langle Bf\rangle$ to vary between epochs.

{Typical $3\sigma$ uncertainties for individual epochs are on the order of $0.3$~kG for $\langle Bf\rangle$, and 30~K for $T_{eff}$, for all stars in the sample. These uncertainties do not include systematics, which previous work has found to be of order $0.1$~dex in logg, 75~K in $T_{eff}$, and 0.25~kG in $\langle Bf\rangle$ \citep{lopez2021}. However, these are not expected to affect our main results, as they rely entirely on a comparative analysis of magnetic fields at different epochs and not on absolute calibration.} 

{We note that the magnetic field distributions on young stars are likely substantially more complex than the single-component magnetic model adopted in this work. Previous studies have shown that multi-component magnetic models can produce systematically different mean field strengths by accounting for a distribution of surface field values and filling factors \citep{yang_magnetic_2008, lavail_characterising_2019}. In the present analysis, the inferred $\langle Bf \rangle$ values should therefore be interpreted as effective average magnetic field strengths rather than unique physical decompositions of the stellar surface magnetic topology.}

{Our adoption of a single-component magnetic model is motivated primarily by the goals of this work and the limitations imposed by the data. Given the substantial rotational broadening present in many targets, together with degeneracies between magnetic broadening, temperature, and veiling, more complex multi-component models would significantly increase parameter degeneracy. While the absolute calibration of the inferred field strengths may therefore depend somewhat on the adopted parameterization, the relative epoch-to-epoch magnetic variability presented here is expected to be robust.}
\section{Results}

\begin{figure}[htb!]
    \centering
    \includegraphics[width=1\linewidth]{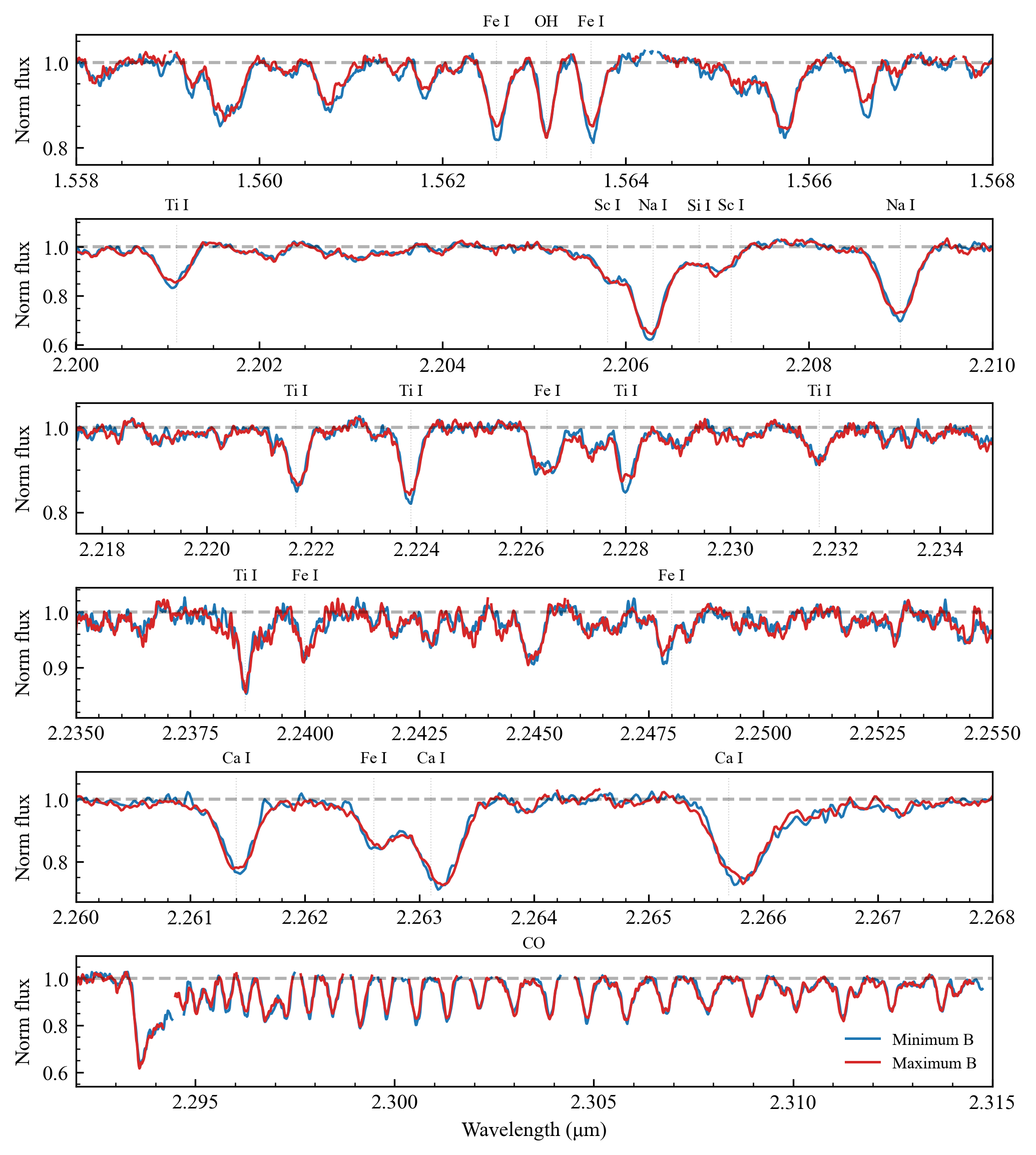}
    \caption{{Comparison between the epochs of minimum and maximum measured surface magnetic field strength in V827~Tau. The blue spectrum corresponds to the minimum-field epoch ($\langle Bf \rangle = 2.15$ kG; JD 2457717.78), while the red spectrum corresponds to the maximum-field epoch ($\langle Bf \rangle = 2.88$ kG; JD 2457782.64).}}
\label{fig:min_vs_max_B}
\end{figure}

{Figure~\ref{fig:min_vs_max_B} compares two epochs of V827~Tau corresponding to the minimum and maximum inferred surface magnetic field strengths in our analysis. The spectra display signs indicative of a different magnetic field strength: the TiI line near 2.228 $\mu$m becomes visibly split during the high-field epoch, while remaining unsplit during the low-field epoch. This transition is consistent with expectations from synthetic spectra \cite[See e.g., Figure~2 of ][]{lopez2021}, which predict the onset of clear Zeeman splitting between approximately 2 and 3 kG. Similarly, the CaI and FeI lines display substantial broadening in the high-field spectrum. The NaI doublet also develops a noticeably flatter line core at higher field strength, again matching the morphology predicted by the magnetic models as the field increases from $\sim$2 to $\sim$3 kG. In contrast, several magnetically insensitive features remain nearly unchanged between epochs. In particular, the OH line in the H-band shows little to no variability despite being highly temperature sensitive. By contrast, the FeI lines around the OH line show   broadening consistent with this interpretation. This is important because the two epochs differ in inferred effective temperature by only $\sim$80 K, indicating that the observed spectral changes cannot be primarily attributed to thermal variability. Instead, the selective enhancement of Zeeman-sensitive lines, combined with the stability of magnetically insensitive features, strongly supports the interpretation that the spectral variability is driven by genuine changes in the surface-averaged magnetic field strength.}

\subsection{Magnetic Variability}\label{sec:mag_var}
\begin{figure*}[htb!]
    \centering
    \includegraphics[width=1\linewidth]{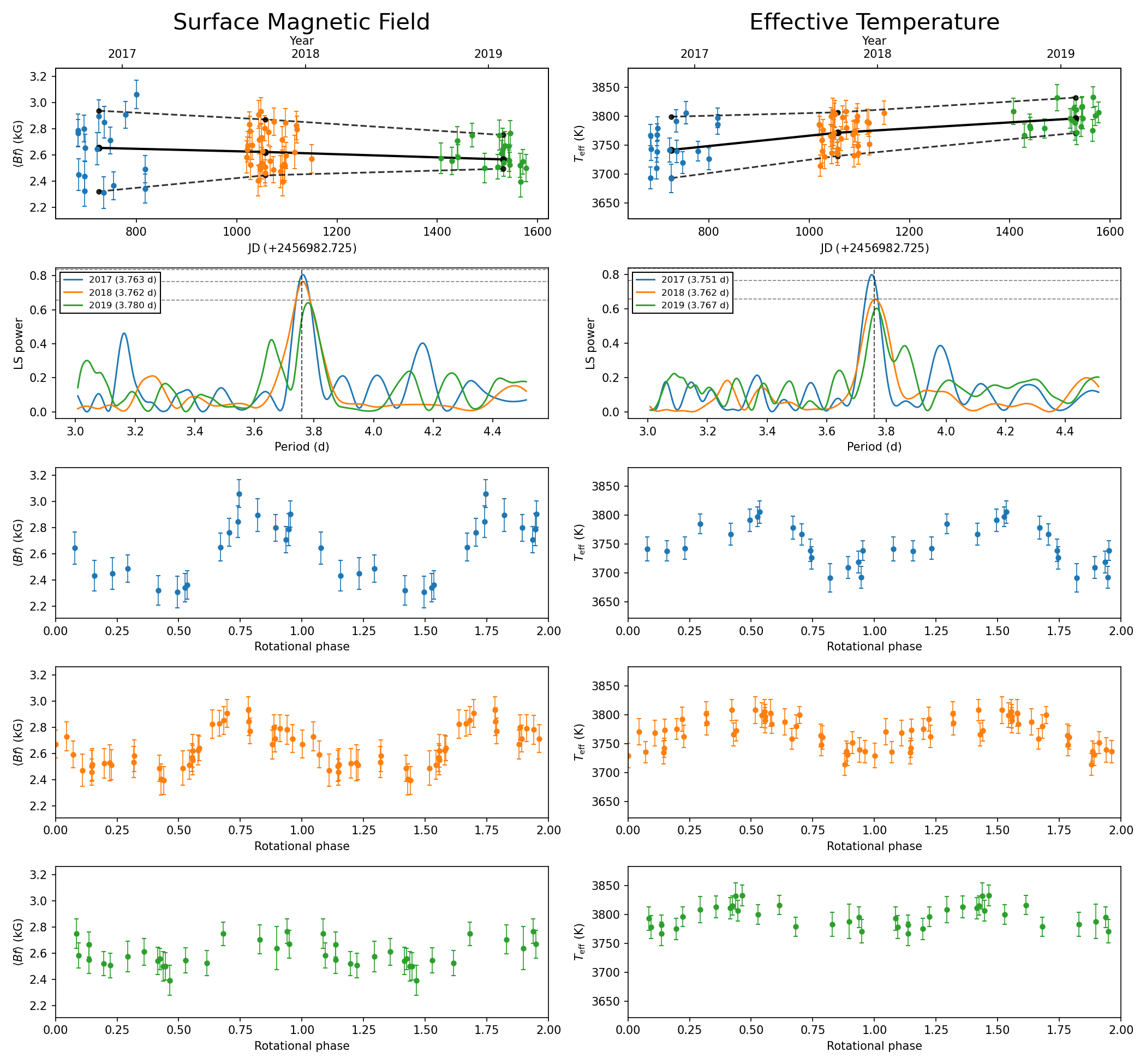}
    \caption{{Time- and phase–resolved magnetic and thermal variability of V827 Tau.} The left column shows the evolution of the mean surface magnetic field strength, $\langle Bf \rangle$, and the right column shows the corresponding best-fit effective temperature, $T_{\mathrm{eff}}$. \textit{Top row:} individual epoch measurements as a function of time, expressed as $\mathrm{JD} - \mathrm{JD}_0$, with points colored by year and overplotted median, 5th, and 95th percentiles for each observing season. \textit{Second row:} Lomb–Scargle periodograms computed independently for each year, with vertical dashed lines indicating the adopted rotation period and horizontal dashed lines marking false alarm probability levels of 0.1, 0.01, and 0.001. \textit{Lower rows:} phase-folded measurements for each year shown separately. Uncertainties throughout correspond to $3\sigma$ posterior estimates.}
\label{fig:v827tau}
\end{figure*}

In Figure~\ref{fig:v827tau} we present the results of our two-step fitting procedure for V827~Tau, based on spectroscopic observations obtained over three observing seasons (2017–2019). The top row of Figure~\ref{fig:v827tau} shows the best-fit effective temperature and mean magnetic field strength (at fixed $\log g$ and $v\sin i$) as a function of Julian Date (JD), together with the 5th, 50th, and 95th percentiles at each epoch, highlighting long-term trends. The second row shows Lomb–Scargle periodograms \citep{Press1989} (computed using the implementation in \texttt{astropy} \citep{astropy2013}) for each observing season using the time series of $\langle Bf \rangle$ and $T_{\mathrm{eff}}$ with Horizontal dashed lines indicate false alarm probability levels of 0.001, 0.01, and 0.1, while a vertical dashed line marks the adopted rotation period from the literature. We were able to recover the rotation period for V827~Tau in 2017 and 2018 in both temperature and $\langle Bf \rangle$ space, but not in 2019, due to this year lacking evidence of rotation variability. 

Over this three-year baseline, the median magnetic field strength remains relatively stable, with values of 2.65~kG in 2017, 2.62~kG in 2018, and 2.56~kG in 2019, in close agreement with the measurements of \citet{lopez2021} who infer a magnetic field strength of 2.42~kG for V827~Tau. The peak-to-peak magnetic variability, defined here as the difference between the 95th and 5th percentiles, evolves significantly with time, decreasing from 0.62~kG in 2017 to 0.26~kG in 2019. In contrast, the median effective temperature exhibits a gradual increase of approximately 50~K from 3742~K in 2017 to 3795~K in 2019. The temperature variability decreased over these three years, from $\Delta T = 106$~K in 2017 to $\Delta T = 61$~K in 2019. The lower panels of Figure~\ref{fig:v827tau} show the phase-folded magnetic field and temperature measurements for each individual year. Moreover, it seems that magnetic variability and temperature variability are anticorrelated, i.e., phases of larger magnetic field strength $\langle Bf \rangle$ correspond to phases of lower effective temperature. We will return to this point in Section~\ref{sec:corr}. 

Across the entire sample, we find evidence for correlated magnetic variability in six out of nine stars (V827~Tau, V830~Tau, L1551-51, CI~Tau, LkCa~15, and AA~Tau), for which we were able to recover the known rotation period in at least one observing season. For the remaining three stars (Hubble~4, GI~Tau, and IQ~Tau), we are unable to detect any variability or rotation period. For these, the observed scatter in $T_{eff}$ and $\langle Bf \rangle$ is comparable to the expected uncertainties, and no clear phase-dependent rotational modulation is detected. This suggests that either the intrinsic variability is weak or that the magnetic and thermal inhomogeneities are distributed in a manner that does not produce a measurable rotational signal in our method.

We find clear differences in the amplitude of the variability and temperature evolution across the sample (See Table~\ref{tab:stars_var}). The largest variations in the magnetic field strengths are observed in AA~Tau and L1551-51, with $\Delta \langle Bf \rangle \sim 0.6$–0.9~kG, as measured from the 5th-95th percentiles in $\langle Bf \rangle$, $\Delta \langle Bf \rangle$,  accompanied by substantial temperature variations of $\Delta T_{\rm eff} \sim 70$–115~K. CI~Tau and V827~Tau exhibit intermediate amplitudes, with both magnetic and thermal variability decreasing over time. In contrast, V1023~Tau (Hubble~4), GI~Tau, and IQ~Tau show consistently low amplitudes in both quantities, with $\Delta \langle Bf \rangle \lesssim 0.2$–0.3 kG and $\Delta T_{\rm eff} \lesssim 50$~K, indicating either intrinsically weak variability or variability below our detection threshold. The largest year-to-year variations in median effective temperature are observed in V827~Tau, CI~Tau, and AA~Tau, with changes of $\sim$50-80~K, while Hubble~4 remains stable at the $\sim$10~K level. This contrast suggests that a subset of stars undergoes genuine secular evolution in surface properties, consistent with long-term magnetic activity cycles or spot evolution, while others remain in a stable state. Therefore, it is not only the mean or rotation-averaged temperature and magnetic field strength that evolve on year-long baselines, but also the amplitude of their rotational-modulated variability, which in some cases weakens or disappears entirely. This behavior suggests changes not just in the global field strength, but in the spatial distribution and contrast of surface magnetic and thermal inhomogeneities.

\begin{deluxetable*}{lccccccccc}
\tablehead{
\colhead{Star} &
\colhead{Year} &
\colhead{$T_{\rm eff}^{5}$} &
\colhead{$T_{\rm eff}^{50}$} &
\colhead{$T_{\rm eff}^{95}$} &
\colhead{$T_{\rm eff}^{5-95}$} &
\colhead{$\langle Bf \rangle^{5}$} &
\colhead{$\langle Bf \rangle^{50}$} &
\colhead{$\langle Bf \rangle^{95}$} &
\colhead{$\langle Bf \rangle^{5-95}$} \\
\colhead{} &
\colhead{} &
\colhead{(K)} &
\colhead{(K)} &
\colhead{(K)} &
\colhead{(K)} &
\colhead{(kG)} &
\colhead{(kG)} &
\colhead{(kG)} &
\colhead{(kG)}
}
\tablecaption{{Year-by-year variability parameters for the sample.}
}
\label{tab:stars_var}
\startdata
AA Tau & 2017 & 3745 & 3775 & 3860 & 115 & 2.21 & 2.36 & 3.06 & 0.85 \\
 & 2019 & 3754 & 3825 & 3862 & 108 & 2.05 & 2.18 & 2.73 & 0.68 \\
CI Tau & 2015 & 4023 & 4062 & 4103 & 80 & 1.73 & 1.86 & 2.24 & 0.51 \\
 & 2017 & 3993 & 4070 & 4107 & 114 & 1.71 & 1.82 & 2.19 & 0.48 \\
 & 2018 & 3943 & 3989 & 4070 & 127 & 1.75 & 1.93 & 2.36 & 0.61 \\
 & 2019 & 3976 & 4030 & 4066 & 90 & 1.87 & 2.13 & 2.22 & 0.36 \\
GI Tau & 2019 & 3747 & 3777 & 3809 & 62 & 2.09 & 2.23 & 2.43 & 0.34 \\
IQ Tau & 2019 & 3746 & 3773 & 3790 & 45 & 1.68 & 1.81 & 1.89 & 0.21 \\
L1551-51 & 2017 & 4178 & 4222 & 4249 & 71 & 2.22 & 2.49 & 2.79 & 0.56 \\
 & 2018 & 4152 & 4192 & 4238 & 85 & 2.49 & 2.69 & 3.09 & 0.60 \\
 & 2019 & 4153 & 4222 & 4252 & 99 & 2.26 & 2.50 & 2.95 & 0.69 \\
LkCa 15 & 2019 & 4217 & 4259 & 4300 & 82 & 1.59 & 1.84 & 1.97 & 0.38 \\
V827 Tau & 2017 & 3693 & 3741 & 3799 & 106 & 2.32 & 2.65 & 2.94 & 0.62 \\
 & 2018 & 3731 & 3771 & 3807 & 75 & 2.44 & 2.62 & 2.87 & 0.43 \\
 & 2019 & 3771 & 3796 & 3832 & 61 & 2.49 & 2.56 & 2.75 & 0.26 \\
V830 Tau & 2016 & 3949 & 3968 & 4023 & 75 & 2.13 & 2.43 & 2.50 & 0.37 \\
 & 2017 & 3979 & 4003 & 4027 & 48 & 2.22 & 2.36 & 2.50 & 0.28 \\
 & 2018 & 3970 & 4002 & 4032 & 61 & 2.10 & 2.42 & 2.56 & 0.46 \\
 & 2019 & 3948 & 3982 & 4005 & 57 & 2.00 & 2.47 & 2.60 & 0.60 \\
V1023 Tau & 2017 & 3858 & 3878 & 3912 & 54 & 2.64 & 2.70 & 2.79 & 0.15 \\
(Hubble 4) & 2018 & 3861 & 3884 & 3904 & 44 & 2.66 & 2.75 & 2.82 & 0.16 \\
 & 2019 & 3860 & 3873 & 3882 & 22 & 2.70 & 2.77 & 2.82 & 0.12 \\
\enddata
\vspace{0.1cm}
For each star and observing year, we report the 5th, 50th (median), and 95th percentiles of the effective temperature, $T_{\rm eff}$, and mean magnetic field strength, $\langle Bf \rangle$. The amplitudes, $T_{\rm eff}^{5-95}$ and $\Delta \langle Bf \rangle^{5-95}$, are defined as the difference between the 95th and 5th percentiles.
\end{deluxetable*}

{We note that between April 2018 and May 2019, IGRINS experienced significant K-band de-focusing, reducing the effective spectral resolution to approximately $R\sim28{,}000$. The magnetic synthetic spectra used throughout this work were generated at the nominal resolution of $R=45,000$. We do not expect this to qualitatively affect our conclusions for several reasons. Even at this reduced resolution, the spectra remain sufficiently resolved to recover Zeeman broadening in the strongly magnetically sensitive lines used throughout this work. To assess the impact of this mismatch, we re-fitted spectra from the defocused period after convolving the model spectra $R=28,000$. The resulting $\langle Bf \rangle$ values differed by $\approx$ 0.05–0.10 kG (well within the formal uncertainties), and the recovered rotational modulation signals remained unchanged. We therefore conclude that the temporary reduction in resolution does not meaningfully affect our measurements or scientific conclusions. 
}

{Hubble~4 has been identified as a spectroscopic binary \citep{Carvalho21_radial_velocity}, which may contribute additional line broadening in its spectrum. In principle, unresolved binarity can mimic or enhance rotational broadening. Based on the available orbital solution, the velocity separation between the two components is substantially smaller than a single IGRINS resolution element. As a result, the system would remain spectroscopically unresolved in our observations and would primarily contribute a small additional broadening term comparable to a modest increase in $v\sin i$. This effect is therefore unlikely to strongly bias the inferred magnetic field strengths. Nevertheless, unresolved multiplicity may contribute additional uncertainty for this star and could explain the absence of a clear magnetic variability detection in Hubble~4. We therefore caution that the inferred field strengths for this source should be interpreted as effective measurements of the unresolved system.}

\subsection{A correlation between magnetic and temperature variability?}\label{sec:corr}
\begin{figure*}[htb!]
    \centering
    \includegraphics[width=1\linewidth]{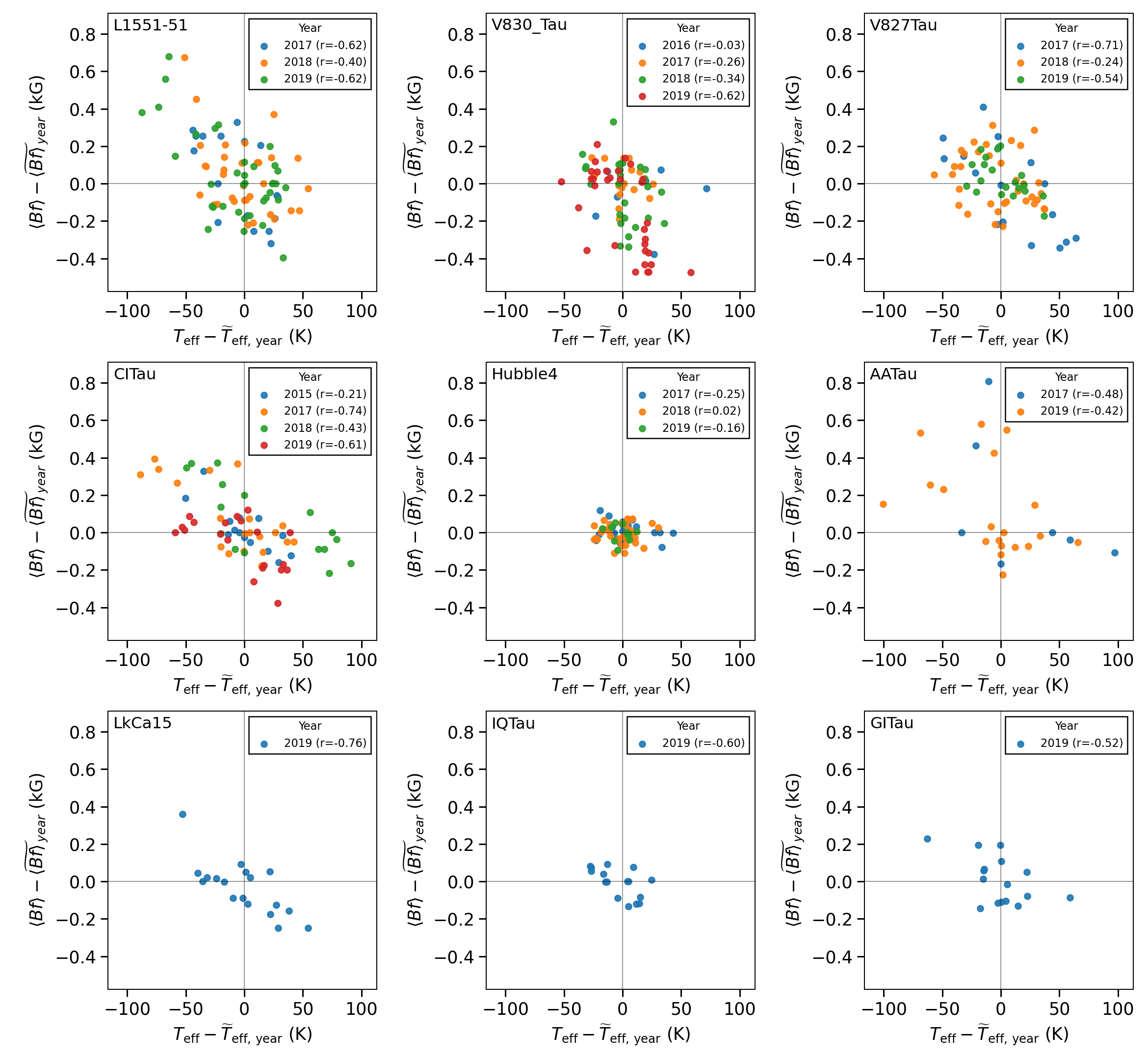}
    \caption{{Scatter Temperature deviations versus magnetic field deviations}. For each star, the quantities plotted are $T_{\rm eff} - \widetilde{T}_{\rm eff,\,year}$ and $\langle Bf \rangle - \widetilde{\langle Bf \rangle}_{\,year}$, where the tilde denotes the median value within each observing year. Points are color-coded by observing year, with the legend indicating the year and the corresponding Pearson correlation coefficient r. Horizontal and vertical lines mark zero deviation in each quantity. Axes limits are uniform across all panels.}
    \label{fig:dBdT}
\end{figure*}

In order to examine the existence of a correlation between magnetic and temperature variability, we infer the coverage of magnetic regions and spots using epoch-by-epoch offsets relative to the median value within each observing season. Specifically, we compute $\Delta \langle Bf \rangle = \langle Bf \rangle - \widetilde{\langle Bf \rangle}_{\,year}$ and $\Delta T_{\rm eff} = T_{\rm eff} - \widetilde{T}_{\rm eff,\,year}$. Figure~\ref{fig:dBdT} shows $\Delta \langle Bf \rangle$ as a function of $\Delta T_{\rm eff}$ for all observations. The resulting distributions exhibit a range of behaviors across stars and observing years. In some cases, correlations are consistently present in all years (e.g., LkCa~15), while in others no correlation is evident (e.g., Hubble~4). Several targets show mixed behavior, with correlations appearing in some years but not others (e.g., L1551-51, V827 Tau). For Hubble~4 in particular, which shows no clear evidence of intrinsic variability, the dispersion in both $\Delta \langle Bf \rangle$ and $\Delta T_{\rm eff}$ is consistent with the expected uncertainties. 

The presence of correlated magnetic and temperature variability in some years, and its apparent absence in others, is consistent with the presence of evolving magnetic activity cycles in pre-main-sequence stars \citep[e.g.,][]{Grankin2008, Grankin2013}, as expected from the dynamo-driven variability observed in other active stars \citep[e.g.,][]{Baliunas1995}. While previous evidence for such cycles has primarily relied on long-term photometric variability and spot evolution, our results provide complementary spectroscopic evidence of the same phenomenon.

\begin{deluxetable*}{lccccccccc}
\tablehead{
\colhead{Star} &
\colhead{Year} &
\colhead{$f_{\rm spot}^{(T),5}$} &
\colhead{$f_{\rm spot}^{(T),50}$} &
\colhead{$f_{\rm spot}^{(T),95}$} &
\colhead{$\Delta f_{\rm spot}^{(T)}$} &
\colhead{$f_{\rm spot}^{(B),5}$} &
\colhead{$f_{\rm spot}^{(B),50}$} &
\colhead{$f_{\rm spot}^{(B),95}$} &
\colhead{$\Delta f_{\rm spot}^{(B)}$}
}
\tablecaption{
{Spot filling factors inferred independently from temperature and magnetic field variability.}
}
\label{tab:fspot_compare}
\startdata
AA Tau & 2017 & 0.73 & 0.70 & 0.59 & 0.14 & 0.79 & 0.88 & 1.27 & 0.48 \\
 & 2019 & 0.72 & 0.63 & 0.59 & 0.14 & 0.70 & 0.78 & 1.08 & 0.38 \\
CI Tau & 2015 & 0.41 & 0.34 & 0.27 & 0.14 & 0.44 & 0.55 & 0.87 & 0.43 \\
 & 2017 & 0.45 & 0.33 & 0.26 & 0.19 & 0.43 & 0.52 & 0.83 & 0.40 \\
 & 2018 & 0.53 & 0.46 & 0.33 & 0.21 & 0.46 & 0.61 & 0.97 & 0.51 \\
 & 2019 & 0.48 & 0.40 & 0.33 & 0.15 & 0.56 & 0.78 & 0.85 & 0.29 \\
\enddata
\vspace{0.1cm}
For each star and observing year, we convert the 5th, 50th, and 95th percentiles of $T_{\rm eff}$ and $\langle Bf \rangle$ into spot filling factors using the fixed two-component parameters from Table~1. The amplitudes are defined as $\Delta f_{\rm spot}=f_{\rm spot}^{5}-f_{\rm spot}^{95}$ for the temperature-based estimates and $\Delta f_{\rm spot}=f_{\rm spot}^{95}-f_{\rm spot}^{5}$ for the magnetic-based estimates, such that both represent the full percentile range in $f_{\rm spot}$.
\end{deluxetable*}

\subsection{Inferring Surface Coverage of Magnetic Regions}\label{sec:surf}
The simultaneous magnetic and thermal variability detected in our sample allows us to test this relationship more directly. In particular, we can use the observed changes in $\langle Bf \rangle$ and $T_{\rm eff}$ to infer the surface coverage of magnetized regions and compare the constraints obtained from the two diagnostics.

\citet{PerezPaolino2025separating} constructed two-component spectral models of spotted stars allowing the temperatures ($T_{spot}$ and $T_{phot}$), magnetic fields ($B_{spot}$ and $B_{phot}$), and filling factors ($f_{phot} = 1 - f_{spot}$) to be free parameters. Applying these models to AA~Tau and CI~Tau, they found $f_{spot}=0.73$, $T_{phot}=4247$~K, $T_{spot}=3499$~K, $B_{phot}=0.80$~kG, and $B_{spot}=2.58$~kG and $f_{spot}=0.40$, $T_{phot}=4249$~K, $T_{spot}=3602$~K, $B_{phot}=1.20$~kG, and $B_{spot}=2.40$~kG, respectively. We use these values to better constrain or understand the relationship or connection between variations in $\langle Bf \rangle$ and $T_{eff}$.


Under a two-component surface model consisting of a quiet photosphere and magnetically active spotted regions, the observed mean magnetic field can be expressed as a surface-weighted average:
\begin{equation}
\langle Bf \rangle = (1 - f_{\rm spot})\, B_{\rm phot} + f_{\rm spot}\, B_{\rm spot},
\end{equation}
where $B_{\rm phot}$ and $B_{\rm spot}$ represent the magnetic field strengths in the photosphere and spot components, respectively. Solving for the filling factor yields:
\begin{equation}
f_{\rm spot}^{(B)} =
\frac{\langle Bf \rangle - B_{\rm phot}}{B_{\rm spot} - B_{\rm phot}}.
\end{equation}

This relation assumes that the stellar surface can be approximated as a linear mixture of two magnetic components. In this framework, $\langle Bf \rangle = B_{\rm phot}$ corresponds to $f_{\rm spot} = 0$, while $\langle Bf \rangle = B_{\rm spot}$ corresponds to $f_{\rm spot} = 1$.

The effective temperature is determined by the total emergent flux, which scales as $T^4$. Assuming again a two-component surface:
\begin{equation}
T_{\rm eff}^4 =
(1 - f_{\rm spot})\, T_{\rm phot}^4 +
f_{\rm spot}\, T_{\rm spot}^4,
\end{equation}
where $T_{\rm phot}$ and $T_{\rm spot}$ are the temperatures of the photospheric and spotted regions, respectively. Rearranging gives:
\begin{equation}
f_{\rm spot}^{(T)} =
\frac{T_{\rm phot}^4 - T_{\rm eff}^4}
     {T_{\rm phot}^4 - T_{\rm spot}^4}.
\end{equation}

This expression provides an alternative estimate of the spot coverage based purely on thermal contrast if we assume $T_{eff}$ is our measured temperature from single-component fits.

For each star and observing year, we compute $f_{\rm spot}$ using both diagnostics by applying the above relations to the 5th, 50th, and 95th percentiles of $\langle Bf \rangle$ and $T_{\rm eff}$. This yields distributions of filling factors:
\begin{equation}
f_{\rm spot}^{(B,\,5,50,95)}\quad \rm{and} \quad
f_{\rm spot}^{(T,\,5,50,95)}.
\end{equation}

\begin{equation}
\Delta f_{\rm spot}^{(B)} =
f_{\rm spot}^{(B,95)} - f_{\rm spot}^{(B,5)},
\end{equation}
\begin{equation}
\Delta f_{\rm spot}^{(T)} =
f_{\rm spot}^{(T,5)} - f_{\rm spot}^{(T,95)},
\end{equation}
where the ordering in the temperature-based amplitude reflects the inverse relationship between $T_{\rm eff}$ and spot coverage.

The comparison between $f_{\rm spot}^{(B)}$ and $f_{\rm spot}^{(T)}$ provides insight into the structure of stellar surfaces. While the temperature-based estimate is primarily sensitive to the coolest regions, the magnetic field-based estimate traces all magnetized surface areas, including those that may not produce a strong thermal contrast.

As a result, we generally expect:
\begin{equation}
f_{\rm spot}^{(B)} \gtrsim f_{\rm spot}^{(T)},
\end{equation}
with equality only in the case where all magnetic regions are sufficiently cool to be detected through their thermal imprint.

We find that the filling factors inferred from the magnetic field, $f_{\rm spot}^{(B)}$, are systematically larger than those inferred from the temperature, $f_{\rm spot}^{(T)}$. The temperature-based estimate traces only regions that are sufficiently cool to produce a measurable reduction in the emergent flux, and is therefore primarily sensitive to  starspots. In contrast, the magnetic field measurement reflects the total unsigned magnetic flux across the stellar surface, including both cool spots and warmer magnetized regions such as plages or intermediate-temperature structures that do not produce a strong thermal contrast. As a result, $f_{\rm spot}^{(B)}$ effectively measures the total area covered by magnetic fields, while $f_{\rm spot}^{(T)}$ traces only the subset of that area associated with the coolest components. The observed inequality $f_{\rm spot}^{(B)} \gtrsim f_{\rm spot}^{(T)}$ therefore provides direct evidence that magnetic regions on young stars extend beyond classical cool spots, occupying a broader fraction of the stellar surface than is apparent from temperature variations alone.

For AA~Tau, we find values of $f_{\rm spot}^{(B,95)}$ exceeding unity. We interpret this as a limitation of the two-temperature, two-magnetic-field-strength framework adopted in \citet{PerezPaolino2025separating}. Starspots are known to host magnetic fields with localized strengths of up to $\sim$8~kG in the Sun, and multi-component modeling of M dwarfs suggests similarly strong fields confined to small surface regions \citep{Shulyak2019}. If the two-component model underestimates the true magnetic field strength within spots, it will systematically overestimate the required surface coverage in order to reproduce the observed $\langle Bf \rangle$. Increasing the assumed spot magnetic field strength to 3~kG, which is typical of magnetic regions \citep{kochukhov_hidden_2020}, and repeating the calculation yields $f_{\rm spot}^{(B,95)} < 1$. This suggests that the apparent inconsistency may be driven by an underestimation of the spot field strength, rather than an intrinsic physical requirement for super-unity filling factors.

\section{Discussion}\label{sec:discussion}
\subsection{Comparison with Previous IGRINS Magnetic Field Measurements}
The detection of coherent rotational modulation and systematic multi-year evolution in $\langle Bf \rangle$ relies on the precision of our magnetic field measurements. Previous work has questioned whether uncertainties at the level of $\sim$0.3 kG (3$\sigma$) are achievable with IGRINS \citep[e.g.,][]{han_magnetic_2023}.

{At the heart of the differences between our results and previous IGRINS magnetic field studies is the choice of spectral regions used during fitting. Both \citet{lopez2021} and \citet{sokal2020} relied primarily on narrow windows in the K band containing a limited number of magnetically sensitive lines. While these regions contain substantial magnetic information, fitting only narrow K-band intervals strengthens degeneracies between parameters such as magnetic field strength, temperature, and surface gravity, particularly because relatively few lines within these windows are insensitive to one or more of the fitted quantities. In practice, this makes it more difficult to independently constrain parameters such as $v\sin i$ and $\langle Bf \rangle$.
}

{Our fitting approach instead simultaneously models both H- and K-band spectral regions. At fixed Landé $g$ factor and magnetic field strength, Zeeman splitting scales as $\lambda^2$, making the H-band region intrinsically less sensitive to magnetic broadening than the K band by a factor of three. This provides an important complementary constraint that helps disentangle temperature, rotational broadening, and magnetic field effects. In particular, we include an H-band window dominated by temperature-sensitive Fe I and OH lines. This window has previously been used to construct an empirical temperature scale for young stars based on the opposite temperature sensitivities of the Fe I and OH lines present in the interval \citep{Tang2024}.
} 

{We therefore expect fitting the K band alone to produce systematically weaker constraints on the surface magnetic field strength. To test this directly, we refit all V830~Tau spectra using only the spectral regions adopted by \citet{lopez2021} and compared the resulting posterior distributions to those obtained from our full H+K fitting procedure. These fits were performed allowing all parameters to vary freely, rather than fixing $T_{\mathrm{eff}}$ and $\log g$, in order to provide a direct comparison between methodologies. We find that restricting the fit to narrow K-band windows increases the inferred magnetic field uncertainties by approximately 30\% for the same spectra. 
}

{\citet{lopez2021} reported differences of order 43–73 K in the inferred temperatures of young stars when comparing fits performed with magnetic models ($\langle Bf \rangle \sim 2$–3 kG) and non-magnetic models. In principle, this raises the possibility that the magnetic variability reported here could instead reflect temperature variability that is being interpreted by the models as changes in magnetic field strength.
}

{To test this, we refit all V827 Tau spectra while fixing $T_{\mathrm{eff}}$, allowing only $\langle Bf \rangle$, $r_H$, and $r_K$ to vary. The resulting best-fit magnetic field strengths differed by less than three percent from those obtained in the fully free fits, demonstrating that our inferred magnetic variability is not strongly driven by temperature degeneracies.}

\subsection{Comparison with ZDI Results}
Four of our nine sample stars, V830~Tau, CI~Tau, AA~Tau, and LkCa~15, have published ZDI magnetograms, providing a useful benchmark for our measurements of the total unsigned magnetic field, $\langle Bf \rangle$. In what follows, we compare the large-scale magnetic topologies and surface brightness distributions inferred from ZDI with the total field strengths and variability amplitudes measured from our Zeeman broadening analysis.

V830~Tau was observed with ESPaDOnS in December 2014 and January 2015 by \citet{donati_magnetic_2015}. Their reconstructed magnetic map indicates that the large-scale field is predominantly poloidal, containing $\sim 90\%$ of the reconstructed magnetic energy, with an average unsigned flux of $\sim 0.3$~kG. The brightness map also includes both cool spots and warm plages, with an overall coverage of $\sim 12\%$ of the surface, roughly equally split between spots and plages. Despite this, the corresponding light curve is relatively flat, indicating that a low photometric amplitude does not imply a low level of surface structure, but rather a more even distribution of brightness features. In contrast, our Zeeman broadening measurements yield mean surface field strengths of $\sim 2.4$--$2.5$~kG between 2016 and 2019, with rotational modulation of up to $\sim 0.6$~kG detected in 2019. The much smaller field strength recovered by ZDI is expected, since ZDI traces only the large-scale component of the magnetic topology, while Zeeman broadening is sensitive to the total unsigned flux, including the dominant small-scale field component unresolved by spectropolarimetry. Our detection of strong magnetic variability in 2019 suggests that since the ZDI observations, the surface distribution of magnetic features evolved significantly, giving rise to rotational modulation in the 0.6~kG range.

LkCa~15 was studied by \citet{Donati2019} using ESPaDOnS observations from November to December 2015. They found that the large-scale magnetic topology is again mostly poloidal, with a strong axisymmetric dipole component of $1.35$~kG and a peak radial field strength reaching $2.2$~kG at the stellar surface. Their magnetic map also includes a large dark photospheric spot at intermediate latitudes, spatially coincident with the strongest radial field region and covering approximately $10$--$20\%$ of the visible hemisphere. In our data, LkCa~15 shows a median $\langle Bf \rangle$ of $1.84$~kG in 2019 and a rotational amplitude of $0.38$~kG, values that fall naturally between the dipolar field strength and the local peak field recovered by ZDI. As in the other stars discussed above, this comparison is consistent with a picture in which the large-scale topology measured by ZDI represents only a fraction of the full surface magnetic structure, while Zeeman broadening captures the total unsigned field averaged over both spotted and non-spotted regions.

CI~Tau has now been observed at multiple epochs with both ESPaDOnS and SPIRou. In their original ESPaDOnS study, \citet{Donati20_magnetic_field} found that CI~Tau hosts a strong large-scale magnetic field, with the radial component reaching $\sim 3.7$~kG in a dark photospheric spot and a dipole component of $\sim 1.7$~kG. More recently, \citet{Donati24_classical_t} monitored CI~Tau with SPIRou in late 2019, late 2020, and late 2022. They found that the large-scale field remained mainly axisymmetric and poloidal, with a dipole component of order $0.8$--$1.1$~kG, while the corresponding longitudinal fields varied on both rotational and multi-year timescales. By adopting a scaling between the large- and small-scale field components, they inferred a small-scale field of order $1.6$--$1.8$~kG in 2019, reaching up to $2.6$~kG by 2022. This is in good agreement with our measurements, which show that the median magnetic field increased from $1.86$~kG in 2015 to $2.13$~kG in 2019, and with those of \citet{sokal2020}. The agreement between the absolute field strengths and their temporal evolution is striking, suggesting that the magnetic field of CI~Tau has undergone genuine long-term evolution over the last decade. Their reconstructed brightness maps also indicate the presence of high-latitude dark spots, which would naturally contribute little to the rotational modulation if located sufficiently close to the pole (for a stellar inclination of i$\sim$50-70$ \degree$). This provides an explanation for why some observing seasons in our data show weaker variability despite still exhibiting strong mean magnetic fields.

For AA~Tau, \citet{Donati10_magnetospheric_accretion} reconstructed the large-scale field from spectropolarimetric observations obtained in December 2008 and January 2009. They found that AA~Tau hosts a strong $2$--$3$~kG dipole tilted by $\sim 20^\circ$ with respect to the rotation axis, and that the magnetic poles coincide with large cool spots at the photospheric level. Their brightness maps show a prominent spotted region centered at intermediate to high latitudes, covering roughly $10\%$ of the stellar surface. Our measurements yield median total field strengths of $2.18$--$2.36$~kG in 2017 and 2019, in good agreement with the strong dipolar component inferred from ZDI, but with much larger filling factors inferred from the magnetic diagnostics. 

For AA Tau and CI Tau, our estimated $\Delta f_{spot}^B$ and $\Delta f_{spot}^T$ of order 0.2 and 0.4 are reasonable when we consider the large spots and magnetic regions recovered by ZDI covering 20-40\% of the surface as the main drivers of variability. This suggests that the remainder of the stellar surface, the missing cold starspots, and the missing magnetic regions, are small and distributed throughout the stellar surface in a way that does not contribute to the overall variability.

Taken together, these comparisons illustrate both the complementarity and the limitations of the two techniques. ZDI provides access to the geometry and topology of the large-scale field, as well as the locations of the dominant cool spots and plages, but it remains insensitive to much of the small-scale mixed-polarity flux and smaller spots. Zeeman broadening, by contrast, cannot recover topology directly, but it does provide a measure of the total unsigned magnetic field and its temporal variability. The broad consistency between the long-term trends inferred from both methods, particularly for CI~Tau, supports the reality of the secular evolution we detect. At the same time, the systematically larger field strengths and filling factors inferred from Zeeman broadening show that most of the magnetic flux on these stars resides outside the large-scale structures recovered by ZDI.

\subsection{Magnetic Variability Measurements of Older Stars Compared to YSOs}
\citet{lavail_characterising_2019} performed multi-epoch Zeeman fitting for a sample of eight T Tauri stars and reported no evidence of strong rotational modulation, with a mean peak-to-peak variation of $\sim$0.3 kG. Their dataset, however, consists of a relatively small number of observations (32 spectra across 8 stars) with limited phase coverage. Notably, the reported amplitude of variability is comparable to the values we find here, suggesting that low-amplitude modulation may be present but difficult to robustly detect with their sampling. Additionally, their CRIRES observations, despite higher spectral resolution (R$\sim$100,000), cover a much narrower K-band wavelength range, limiting the number of usable magnetically sensitive and insensitive lines compared to the IGRINS data presented here and leading to a lower sensitivity. 

Our results are qualitatively consistent with the findings of \citet{Cristofari2025}, who report an anti-correlation between magnetic field strength and differential temperature (d$Temp$) measured from high-resolution spectra for a sample of M-dwarfs. While their d$Temp$ diagnostic is derived from a linearized projection of spectral variability from a reference temperature, and therefore differs from our directly inferred $T_{\rm eff}$ values, both quantities trace the rotational modulation of magnetic surface features. In particular, cool, magnetically active regions such as starspots are expected to simultaneously increase the disk-integrated magnetic field strength and decrease the apparent temperature of the stellar photosphere when they appear on the visible stellar hemisphere. As a result, both d$Temp$ and $\Delta T_{\rm eff}$ act as proxies for variability of the spot coverage on the observable hemisphere, and are therefore expected to correlate with $\langle Bf \rangle$. The agreement between these independent approaches strengthens the interpretation that the observed variability is driven by magnetic inhomogeneities on the stellar surfaces rather than stochastic noise or degeneracies among the fitted parameters. However, the presence of magnetic regions associated with regions that do not show temperature variations with the unspotted photosphere, or are hotter (e.g., plages) is possible.

More recently, \citet{Hahlin26_activity_correlation} monitored four young solar analogs, finding clear rotational modulation of the total unsigned magnetic field in one star and tentative evidence in another. They further report positive correlations between the total unsigned magnetic field strength and common activity indices, while no comparable relationship is found for the large-scale field inferred from ZDI. Since these activity diagnostics are expected to be more sensitive to the unsigned surface magnetic flux associated with small-scale fields, their results suggest that small-scale magnetic structures play a dominant role in linking magnetic variability to stellar activity. Taken together, these results suggest that rotationally modulated magnetic variability is widespread among young and fully convective stars.

\section{Conclusions}\label{sec:conclusion}

We have conducted an analysis of multi-year, high-cadence IGRINS spectroscopy for nine T~Tauri stars, comprising 489 near-infrared spectra from the RRISA archive. By fitting each epoch with synthetic magnetic spectra, we recovered mean surface magnetic field strengths and effective temperatures, allowing us to trace magnetic and thermal variability on both rotational and multi-year timescales. Six of the nine stars exhibit phase-coherent modulation in $\langle Bf\rangle$ and $T_{\rm eff}$, and we recovered the known rotation period in at least one season for all six. Both the mean field strengths and temperatures evolve from year to year, and the amplitudes of rotational modulation vary significantly, sometimes decreasing or vanishing entirely. These trends indicate that variability arises not only from changes in total magnetic flux but also from evolving distributions of magnetic and thermal surface features.

For AA~Tau and CI~Tau, we used two-component spot models to infer surface coverage fractions independently from the magnetic and thermal diagnostics. In both stars, the magnetic filling factors exceed those dervied for starspots, implying that the total magnetized area is larger than the area occupied by the coolest starspots and includes warmer magnetized regions. Comparisons with published Zeeman–Doppler imaging suggest that the large-scale field maps recover only a small fraction of the total magnetic flux; our Zeeman broadening measurements reveal mean surface fields of $\sim 1$–3~kG, with filling factors larger than those inferred from ZDI. Together, these results demonstrate that pre-main-sequence magnetic fields are structured, vary on rotational timescales, and evolve on multi-year baselines, and that small-scale fields dominate the surface flux. Continued monitoring with high-resolution near-infrared spectroscopy, coupled with contemporaneous spectropolarimetry and photometry, will be essential to map the full spectrum of magnetic variability in young stars. Such knowledge will enable us to quantify how magnetic activity modulates the high-energy environments that shape nascent planetary atmospheres.

\begin{acknowledgments}
{We wish to thank Erica Sawczynec, Greg Mace, Dan Jaffe, and Adolfo Carvalho for insightful conversations and suggestions that strengthened this manuscript. We are grateful to the many IGRINS Team members, PIs, and observatory staff whose efforts and time spent designing, proposing, maintaining, observing, and reducing the data from IGRINS made the RRISA archive possible. Their time investment and expertise created the dataset that enabled this study. This paper makes use of the stellar research environment created by the University of Texas at Austin Infrared Spectroscopy Group, including the high-resolution near-infrared spectrometer IGRINS \citep{Park2014, Mace16_300}, the IGRINS data analysis pipeline  \citep[IGRINS PLP; ][]{kaplan2024}, and the polarized stellar radiative transfer code MOOGSTOKES \citep{Deen2013}. The Immersion Grating Infrared Spectrometer (IGRINS) was developed under a collaboration between the University of Texas at Austin and the Korea Astronomy and Space Science Institute (KASI) with the financial support of the US National Science Foundation under grants AST-1229522, AST-1702267, and AST-1908892, McDonald Observatory of the University of Texas at Austin, the Korean GMT Project of KASI, and the Mt. Cuba Astronomical Foundation. The RRISA is maintained by the IGRINS Team with support from McDonald Observatory of the University of Texas at Austin and the US National Science Foundation under grant AST-1908892. 
}
\end{acknowledgments}

\pagebreak
\bibliography{spots_working.bib}{}
\bibliographystyle{aasjournalv7}

\setcounter{figure}{0}
\renewcommand{\thefigure}{A\arabic{figure}}
\appendix

\begin{figure*}
    \centering
    \includegraphics[width=1\linewidth]{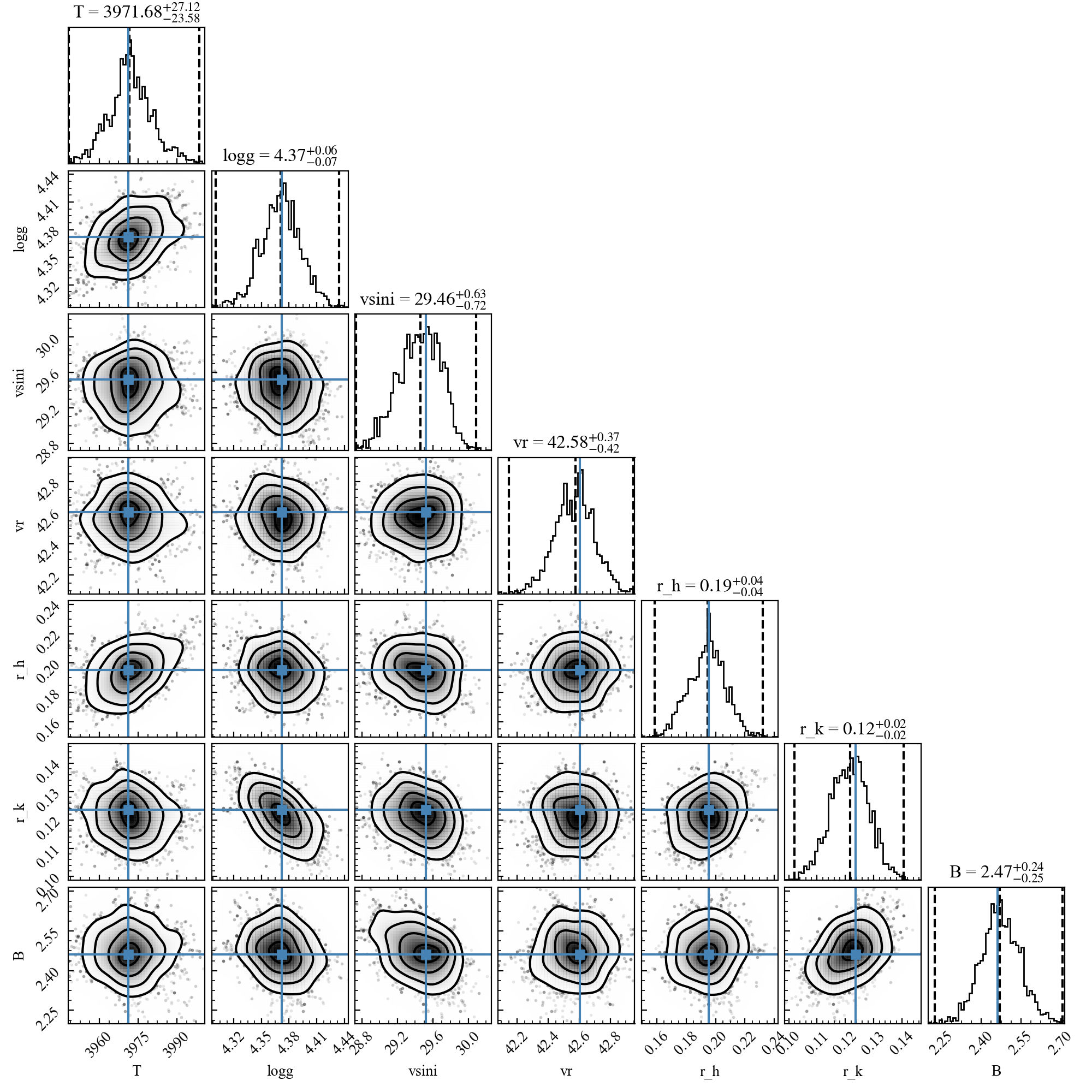}
    \caption{{Example corner plot for V830~Tau with $3\sigma$ uncertainties}}
    \label{fig:corner}
\end{figure*}
\end{document}